\documentclass[prc,preprintnumbers,onecolumn]{revtex4}
\usepackage{graphicx}
\usepackage{amssymb}
\usepackage{amsmath}

\begin{document}

\title{Derivation of capture and reaction cross sections from experimental  quasi-elastic and  elastic
backscattering probabilities}
\author{V.V.Sargsyan$^{1,2}$, G.G.Adamian$^{1}$, N.V.Antonenko$^1$,   and   P.R.S.Gomes$^3$
}
\affiliation{$^{1}$Joint Institute for Nuclear Research, 141980 Dubna, Russia\\
$^{2}$International Center for Advanced Studies, Yerevan State University, 0025 Yerevan, Armenia\\
$^{3}$Instituto de Fisica, Universidade Federal Fluminense, Av. Litor\^anea, s/n, Niter\'oi, R.J. 24210-340, Brazil
}
\date{\today}



\begin{abstract}
  We suggest simple and useful
methods to extract reaction and capture (fusion) cross sections from the experimental
elastic and quasi-elastic backscattering data.
\end{abstract}

\maketitle
\section{Introduction}
\label{intro}
The direct measurement
of the  reaction or  capture (fusion) cross section is   a difficult task
since it would require the measurement of individual cross sections of many reaction
channels, and most of them could be reached only by specific experiments.
This would require different experimental set-ups not always available at the
same laboratory and, consequently, such direct measurements would demand a
large amount of beam time and would take probably some years to be
reached. Because of that,
the measurements of elastic scattering angular distributions that
cover full angular ranges and optical model analysis have been used for
the determination of reaction cross sections. This traditional method
consists in deriving the parameters of the complex optical potentials which
fit the experimental elastic scattering angular distributions and then of
deriving the reaction cross sections predicted by these potentials.
 Even so, both the experimental part
and the analysis of this latter method are not so simple.
In the present
work we present a much simpler methods to determine
reaction and capture (fusion) cross sections.
They consist  of measuring only elastic or quasi-elastic scattering at one
backward angle, and from that, the extraction of the reaction or capture cross sections
can   easily be performed.

\section{Relationship  between capture and quasi-elastic backscattering,
and relationship between reaction and elastic backscattering}
\label{sec-1}
From the conservation of the total reaction flux one can write \cite{Sargsyan13b,Sargsyan13c,Sargsyan13a,Canto06} the expressions
\begin{equation}
P_{el}(E_{\mathrm{c.m.}},J)+P_{R}(E_{\mathrm{c.m.}},J)=1
\label{el_eq}
\end{equation}
and
\begin{equation}
P_{qe}(E_{\mathrm{c.m.}},J)+P_{cap}(E_{\mathrm{c.m.}},J)+P_{BU}(E_{\mathrm{c.m.}},J)+P_{DIC}(E_{\mathrm{c.m.}},J)=1.
\label{cn_eq}
\end{equation}
Quasi-elastic scattering probability
\begin{equation}
P_{qe}(E_{\mathrm{c.m.}},J)=P_{el}(E_{\mathrm{c.m.}},J)+P_{in}(E_{\mathrm{c.m.}},J)+P_{tr}(E_{\mathrm{c.m.}},J)
\end{equation}
is defined as the sum of elastic scattering $P_{el}$,
inelastic excitations $P_{in}$ and a few nucleon transfer $P_{tr}$ probabilities.  The  reaction
probability may be written as
\begin{equation}
P_{R}(E_{\mathrm{c.m.}},J)=P_{in}(E_{\mathrm{c.m.}},J)+P_{tr}(E_{\mathrm{c.m.%
}},J)+P_{cap}(E_{\mathrm{c.m.}},J)+P_{BU}(E_{\mathrm{c.m.}},J)+P_{DIC}(E_{%
\mathrm{c.m.}},J),
\end{equation}%
where
$P_{cap}$ is the capture probability
(sum of  evaporation-residue formation, fusion-fission, and
quasi-fission probabilities or sum of  fusion and quasi-fission probabilities),
$P_{DIC}$ is the deep inelastic collision
probability, and $P_{BU}$ is the breakup probability, important particularly
when weakly bound nuclei are involved in the reaction \cite{Canto06}. Note
that the deep inelastic collision process is only important at large
energies above the Coulomb barrier.
%
Here  we neglect the deep inelastic
collision process, since we are concerned with low energies. Thus, one can extract
the reaction
\begin{equation}
P^{ex}_{R}(E_{\mathrm{c.m.}},J=0)=1-P_{el}(E_{\mathrm{c.m.}},J=0)
\label{1cn_eq}
\end{equation}
and
capture
\begin{equation}
P^{ex}_{cap}(E_{\rm c.m.},J=0)=1-[P_{qe}(E_{\rm c.m.},J=0)+P_{BU}(E_{\rm c.m.},J=0)]
\end{equation}
probabilities at $J=0$
from the
experimental elastic backscattering probability $P_{el}(E_{\mathrm{c.m.}},J=0)$
and   quasi-elastic backscattering  probability $P_{qe}(E_{\mathrm{c.m.}},J=0)$ plus
breakup probability $P_{BU}(E_{\rm c.m.},J=0)$ at backward angle, respectively.
Here, the elastic or quasi-elastic scattering or breakup probability~\cite%
{Canto06,Timmers,Timmers1,Timmers2,Zhang}
\begin{equation}
P_{el,qe,BU}(E_{\mathrm{c.m.}},J=0)=d\sigma _{el,qe,BU}/d\sigma _{Ru}
\end{equation}
for angular momentum $J=0$ is given by the ratio of the elastic or quasi-elastic scattering or breakup
differential cross section and Rutherford differential cross section at  180
degrees.
Furthermore, one can approximate the $J$ dependence of the reaction $P_{R}(E_{\mathrm{c.m.}},J)$
and capture $P_{cap}(E_{\mathrm{c.m.}},J)$
probabilities at a given bombarding energy $E_{\mathrm{c.m.}%
} $ by shifting the energy \cite{Sargsyan13b,Sargsyan13c}:
\begin{equation}
P_{R}(E_{\mathrm{c.m.}},J)\approx P^{ex}_{R}(E_{\mathrm{c.m.}}-\frac{\hbar
^{2}\Lambda }{2\mu R_{b}^{2}}-\frac{\hbar ^{4}\Lambda ^{2}}{2\mu ^{3}\omega_{b}^{2}R_{b}^{6}},J=0)
\label{2cn_eq}
\end{equation}
and
\begin{eqnarray}
P_{cap}(E_{\rm c.m.},J)\approx P^{ex}_{cap}(E_{\rm c.m.}-
\frac{\hbar^2\Lambda}{2\mu R_b^2}-\frac{\hbar^4\Lambda^2}{2\mu^3\omega_{b}^{2}R_b^6},J=0),
\end{eqnarray}
where $\Lambda =J(J+1)$, $R_{b}=R_{b}(J=0)$ is the position of the Coulomb
barrier at $J=0$, $\mu =m_{0}A_{1}A_{2}/(A_{1}+A_{2})$ is the reduced mass ($%
m_{0}$ is the nucleon mass), and $\omega _{b}$ is the curvature of the
s-wave potential barrier.  Here we used the expansion of the height $V_b(J)$ of the Coulomb barrier
up to second order in $\Lambda$ \cite{Sargsyan13b,Sargsyan13c}:
$$V_b(J)=V_b(J=0)+\frac{\hbar^2\Lambda}{2\mu R_b^2}+\frac{\hbar^4\Lambda^2}{2\mu^3\omega_{b}^{2}R_b^6}.$$
Employing formulas for the reaction
\begin{eqnarray}
\sigma_{R}(E_{\rm c.m.})=
\pi\lambdabar^2
\sum_{J=0}^{\infty}(2J+1)P^{ex}_{R}(E_{\rm c.m.}-
\frac{\hbar^2\Lambda}{2\mu R_b^2}-\frac{\hbar^4\Lambda^2}{2\mu^3\omega_{b}^{2}R_b^6},J=0)
\label{1aaa_eq}
\end{eqnarray}
and capture
\begin{eqnarray}
\sigma_{cap}(E_{\rm c.m.})=
\pi\lambdabar^2
\sum_{J=0}^{J_{cr}}(2J+1)P^{ex}_{cap}(E_{\rm c.m.}-
\frac{\hbar^2\Lambda}{2\mu R_b^2}-\frac{\hbar^4\Lambda^2}{2\mu^3\omega_{b}^{2}R_b^6},J=0)
\label{1ab_eq}
\end{eqnarray}
cross sections,
converting the sum over the partial waves $J$ into an integral, and
expressing $J$ by the variable $E=E_{\mathrm{c.m.}}-\frac{\hbar ^{2}\Lambda
}{2\mu R_{b}^{2}}$, we obtain the following simple expressions \cite{Sargsyan13b,Sargsyan13c}:
\begin{equation}
\sigma _{R}(E_{\mathrm{c.m.}})=\frac{\pi R_{b}^{2}}{E_{\mathrm{c.m.}}}%
\int_{0}^{E_{\mathrm{c.m.}}}dEP^{ex}_{R}(E,J=0)[1-\frac{%
4(E_{\mathrm{c.m.}}-E)}{\mu \omega _{b}^{2}R_{b}^{2}}]
\label{3rcn_eq}
\end{equation}
and
\begin{eqnarray}
\sigma_{cap}(E_{\rm c.m.})
=\frac{\pi R_b^2}{E_{\rm c.m.}}
\int_{E_{\rm c.m.}-\frac{\hbar^2\Lambda_{cr}}{2\mu R_b^2}}^{E_{\rm c.m.}}dEP^{ex}_{cap}(E,J=0)[1-\frac{4(E_{\rm c.m.}-E)}{\mu\omega_b^2  R_b^2}],
\label{3capcn_eq}
\end{eqnarray}
where $\lambdabar^2=\hbar^2/(2\mu E_{\rm c.m.})$ is the reduced de Broglie wavelength,
 $\Lambda_{cr}=J_{cr}(J_{cr}+1)$,
and $J=J_{cr}$ is the critical angular momentum.
For values $J$ greater than $J_{cr}$, the potential pocket in the
nucleus-nucleus interaction potential vanishes and the capture is not occur.
To calculate the critical angular momentum $J_{cr}$ and
 the position $R_{b}$
of the Coulomb barrier,
we use the nucleus-nucleus interaction potential $V(R,J)$ of  Ref.~\cite{Pot}.
For the nuclear part of the nucleus-nucleus potential, the double-folding formalism with
the Skyrme-type density-dependent effective nucleon-nucleon interaction is employed~\cite{Pot}.

For the systems with $Z_1\times Z_2 < 2000$ ($Z_{1,2}$ are the atomic numbers of interacting nuclei),
the critical angular momentum $J_{cr}$ is large enough
and Eq. (\ref{3capcn_eq}) can be approximated with  good accuracy as:
\begin{eqnarray}
\sigma_{cap}(E_{\rm c.m.})
\approx \frac{\pi R_b^2}{E_{\rm c.m.}}
\int_{0}^{E_{\rm c.m.}}dEP^{ex}_{cap}(E,J=0)[1-\frac{4(E_{\rm c.m.}-E)}{\mu\omega_b^2  R_b^2}].
\label{3cx_eq}
\end{eqnarray}

The formula (\ref{3rcn_eq}) [(\ref{3capcn_eq})] relates the reaction [capture] cross section with elastic [quasi-elastic]
scattering excitation function at a backward angle. By using the experimental
$P_{el}(E_{\mathrm{c.m.}},J=0)$ [$P_{qe}(E_{\mathrm{c.m.}},J=0)$+$P_{BU}(E_{\mathrm{c.m.}},J=0)$] and Eq. (\ref{3rcn_eq}) [(\ref{3capcn_eq})],
 one can obtain the reaction [capture] cross sections.

It is important to mention that since the generalized form of the optical
theorem connects the reaction cross section and forward elastic scattering
amplitude\cite{Canto06},  we show that the forward and
backward elastic scattering amplitudes are related to each other.

Using the extracted $\sigma_{cap}$ and the experimental
$P_{qe}$, one can find the average angular momentum
\begin{eqnarray}
<J>
=\frac{\pi R_b^2}{E_{\rm c.m.}\sigma_{cap}(E_{\rm c.m.})}
\int_{E_{\rm c.m.}-\frac{\hbar^2\Lambda_{cr}}{2\mu R_b^2}}^{E_{\rm c.m.}}
&&dEP^{ex}_{cap}(E,J=0)
[1-\frac{5(E_{\rm c.m.}-E)}{\mu\omega_b^2 R_b^2}]\nonumber\\
&&\times [(\frac{2\mu R_b^2}{\hbar^2}(E_{\rm c.m.}-E)+\frac{1}{4})^{1/2}-\frac{1}{2}]
\label{3cxJ_eq}
\end{eqnarray}
and the second moment of the angular momentum
\begin{eqnarray}
<J(J+1)>
=\frac{2\pi\mu R_b^4}{\hbar^2E_{\rm c.m.}\sigma_{cap}(E_{\rm c.m.})}
\int_{E_{\rm c.m.}-\frac{\hbar^2\Lambda_{cr}}{2\mu R_b^2}}^{E_{\rm c.m.}}
&&dEP^{ex}_{cap}(E,J=0)
[1-\frac{6(E_{\rm c.m.}-E)}{\mu\omega_b^2 R_b^2}]\nonumber\\
&&\times [E_{\rm c.m.}-E]
\label{3cxJ2_eq}
\end{eqnarray}
of the captured system \cite{Sargsyan13b}.

\section{Results of calculations}
\label{sec-2}
As the   elastic, quasi-elastic, and breakup  data were not taken at 180
degrees, but rather at backward angles in the range from 150 to 170 degrees,
the corresponding center of mass energies were corrected by the centrifugal
potential at the experimental angle \cite{Timmers}.
\begin{figure}[htb]
\centering
\includegraphics[scale=1]{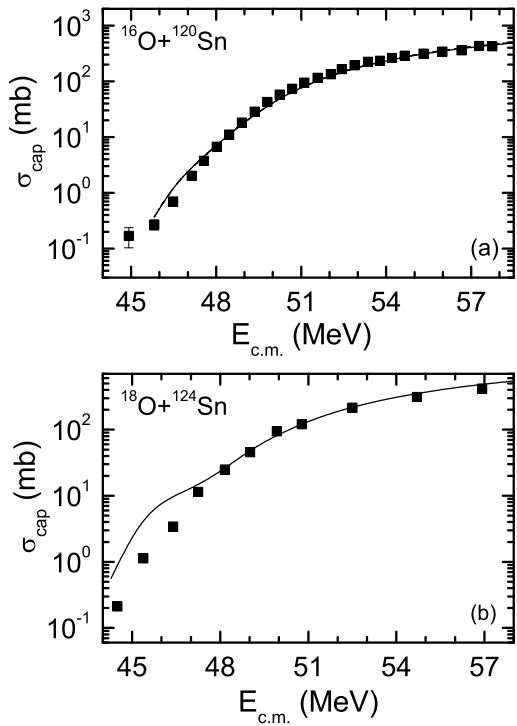}
\caption{
The extracted capture cross sections for the  reactions $^{16}$O + $^{120}$Sn (a) and  $^{18}$O + $^{124}$Sn (b)
by employing
Eq.~(13) (solid line) and Eq.~(14) (dotted line). These lines are almost coincide.
The used experimental quasi-elastic backscattering data are from Ref.~\protect\cite{Sinha}.
The experimental capture (fusion) data (symbols) are from Refs.~\protect\cite{Sinha,JACOBS}.
}
\label{1_fig}
\end{figure}
\begin{figure}[htb]
\centering
\includegraphics[scale=1]{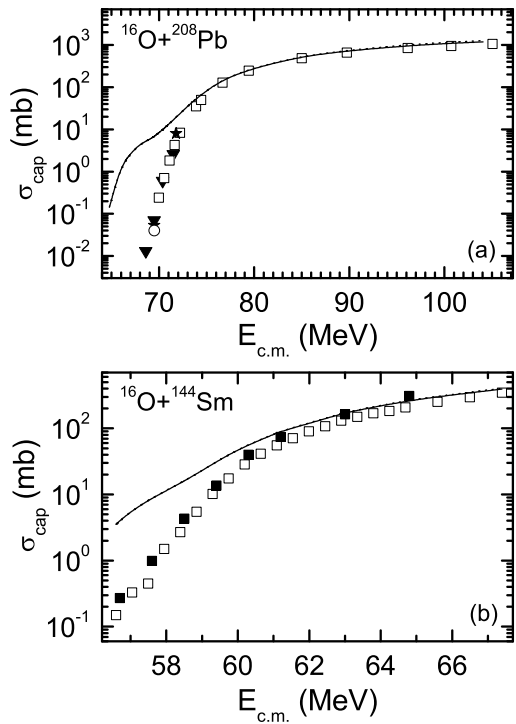}
\caption{The same as in Fig.~1, but  for the reactions
$^{16}$O + $^{208}$Pb(a),$^{144}$Sm(b).
The used  experimental quasi-elastic backscattering  data are from Refs.~\protect\cite{Timmers2,Timmers}.
For the  $^{16}$O + $^{208}$Pb reaction, the experimental capture (fusion) data are from
Refs.~\protect\cite{Pbcap} (open squares),~\protect\cite{Pbcap1} (open circles),~\protect\cite{Pbcap2} (closed stars),
and~\protect\cite{Pbcap3} (closed triangles).
For the $^{16}$O + $^{144}$Sm  reaction, the experimental capture (fusion) data  are
from Refs.~\protect\cite{SmCap1} (closed squares) and~\protect\cite{SmCap2} (open squares).
}
\label{2_fig}
\end{figure}
\subsection{Capture cross sections}
\label{sec-3}
For the verification of our method of the extraction of $\sigma_{cap}$,
first we compare the extracted
capture cross sections with experimental one for the reactions with toughly bound nuclei [$P_{BU}(E_{\mathrm{c.m.}},J=0)=0$].
In Figs.~1 and 2 one can see  good agreement between the extracted and directly
measured capture cross sections for the reactions $^{16}$O + $^{120}$Sn, $^{18}$O + $^{124}$Sn,
$^{16}$O + $^{208}$Pb, and $^{16}$O + $^{144}$Sm at energies above the Coulomb barrier.
The results on the sub-barrier energy region are discussed later on.
To extract the capture cross section, we use both Eq.~(13) (solid lines) and Eq.~(14) (dotted lines).
The used values of critical angular momentum are $J_{cr}$=54, 56, 57, and 62
for the reactions $^{16}$O + $^{120}$Sn, $^{18}$O + $^{124}$Sn,
$^{16}$O + $^{144}$Sm, and $^{16}$O + $^{208}$Pb, respectively.
The difference between the
results of Eqs. (13) and (14)  is less than 5$\%$ at the highest
energies. At low energies,  Eqs.~(13) and (14) lead to the same values
of  $\sigma_{cap}$. The factor $1-\frac{4(E_{\rm c.m.}-E)}{\mu\omega_b^2  R_b^2}$ in  Eqs. (13) and (14)
very weakly influences the results of the  calculations for the systems and energies considered.
Hence, one can say that for the relatively light systems the proposed method of extracting the capture cross section
is model independent (particular, independent on the potential used):
$$\sigma_{cap}(E_{\rm c.m.})
\approx \frac{\pi R_b^2}{E_{\rm c.m.}}
\int_{0}^{E_{\rm c.m.}}dEP^{ex}_{cap}(E,J=0).$$

One can see that the used formulas are suitable not only for  almost spherical nuclei
(Figs. 1 and 2)  but also for the reactions
with strongly deformed target- or projectile-nucleus (Figs. 3 and 4). So,
the deformation effect is effectively contained  in the experimental $P_{qe}$.
$J_{cr}=58$,
 68, 74, and 76 for the reactions  $^{16}$O+$^{154}$Sm,
$^{32}$S+$^{90}$Zr,  $^{32}$S+$^{96}$Zr, and $^{20}$Ne+$^{208}$Pb, respectively.
The results obtained by employing the formula (14) are almost the same and not presented in Figs. 3 and 4.
\begin{figure}[htb]
\centering
\includegraphics[scale=1]{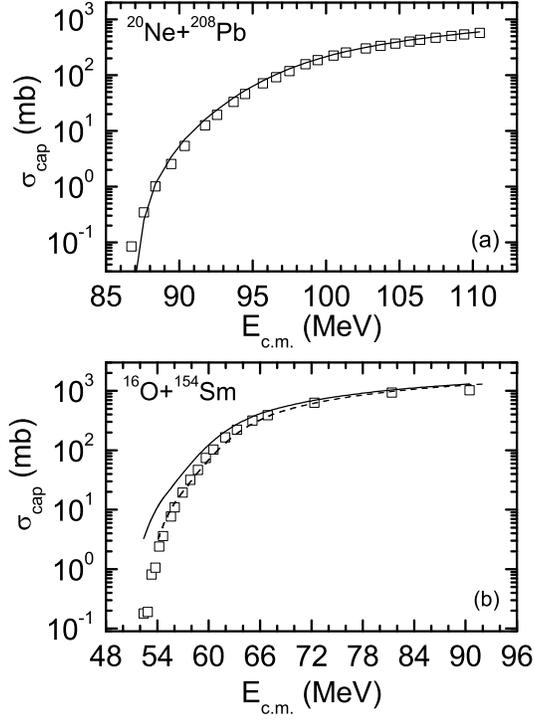}
\caption{
The same as in Fig.~1, but for the  reactions $^{20}$Ne + $^{208}$Pb and $^{16}$O + $^{154}$Sm.
The used  experimental quasi-elastic backscattering  data are from Refs.~\protect\cite{Piasecki,Timmers}.
The experimental capture (fusion) data (symbols) are from Refs.~\protect\cite{SmCap2,Piasecki}.
For the   $^{16}$O + $^{154}$Sm  reaction,
the dashed line is obtained from the shift of the solid line  by  1.7 MeV  to  higher  energies.
}
\label{3_fig}
\end{figure}

\begin{figure}[htb]
\centering
\includegraphics[scale=1]{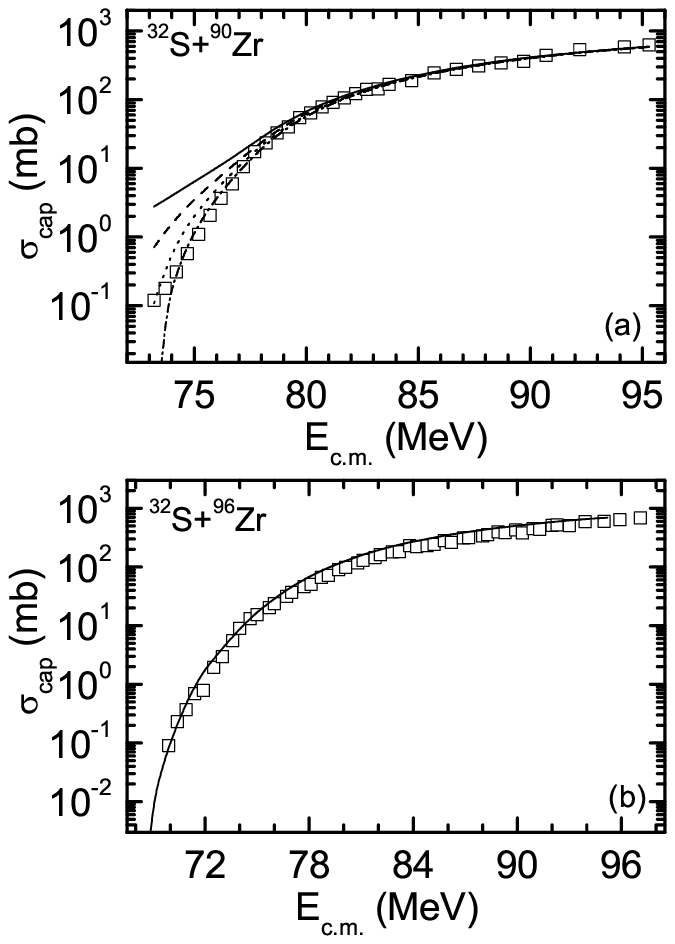}
\caption{
The same as in Fig.~1, but for the  reactions $^{32}$S + $^{90}$Zr (a) and $^{32}$S + $^{96}$Zr (b).
For the  $^{32}$S+$^{90}$Zr   reaction, we show the extracted capture cross sections,
increasing  the experimental $P_{qe}$ by 1\% (dashed line), 2\% (dotted line), and 3\% (dash-dotted line).
The  used experimental quasi-elastic  backscattering data are from Ref.~\protect\cite{Zhang3}.
The experimental capture (fusion) data (symbols) are from Ref.~\protect\cite{ZhangS32Zn9096}.
For the $^{32}$S + $^{96}$Zr  reaction, the energy scale for the extracted capture cross sections
is adjusted to that of the direct measurements.
}
\label{4_fig}
\end{figure}
For the reactions   $^{16}$O+$^{154}$Sm  and  $^{32}$S+$^{96}$Zr,
the extracted capture cross sections are shifted in energy by 1.7 and 1.9 MeV with respect to the measured capture data, respectively.
This could be
the result of different energy calibrations in the experiments on the capture measurement and on the
quasi-elastic scattering. Because of the lack of systematics in these energy shifts, their
origin remains unclear and we adjust  the
Coulomb barriers in the extracted capture cross sections to the values following the experiments.

Note that the extracted
and experimental capture cross sections deviate from each other in the reactions
$^{16}$O+$^{208}$Pb, $^{16}$O+$^{144}$Sm, and  $^{32}$S+$^{90}$Zr at energies below the Coulomb barrier.
Probably this deviation (the mismatch between quasi-elastic backscattering and fusion (capture)
experimental data)
is a reason for the large discrepancies
in the diffuseness parameter extracted from the analyses
of the quasi-elastic scattering and fusion (capture) at deep sub-barrier energies.
One of the possible reasons for the overestimation of the capture cross section from the quasi-elastic data
at sub-barrier energies is
the underestimation of the total reaction differential  cross section taken
as the Rutherford differential cross section.
Indeed, for the  $^{32}$S+$^{90}$Zr   reaction,
the increase of $P_{qe}$ within 2--3\% is needed in order to obtain
the agreement between the extracted and measured
capture cross sections at the sub-barrier energies [Fig.~4(a)].

As seen in Fig.~5,
the extracted capture cross sections $\sigma_{cap}(E_{\rm c.m.})$ (solid line)
for the $^{6}$Li+$^{208}$Pb reaction with weakly bound nucleus [$P_{BU}(E_{\rm c.m.},J=0)\neq 0$] are
rather close to those found in the direct measurements~\cite{Li6Pbcap}
at energies above the Coulomb barrier.
\begin{figure}[htb]
\centering
\includegraphics[scale=0.75]{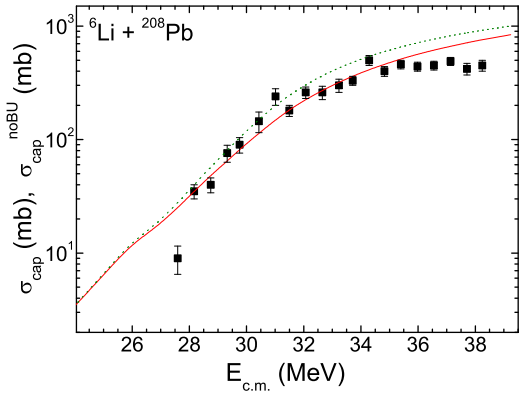}
\caption{
(Color online)
The extracted capture cross sections $\sigma_{cap}(E_{\rm c.m.})$ (solid line)  and
$\sigma^{noBU}_{cap}(E_{\rm c.m.})$  (dotted line)
for the $^{6}$Li+$^{208}$Pb reaction.
The used experimental quasi-elastic  backscattering and quasi-elastic  backscattering  plus breakup at
the backward angle
data are from Ref.~\protect\cite{Li6Pb}.
The experimental capture cross sections (solid squares)  are from Ref.~\protect\cite{Li6Pbcap}.
The energy scale for the extracted capture cross sections
is adjusted to that of the direct measurements.
}
\label{5_fig}
\end{figure}
It appears that at energies near and below the Coulomb barrier
 the extracted $\sigma_{cap}(E_{\rm c.m.})$ deviates from the direct measurements.
It is similarly possible to calculate the capture excitation function
\begin{eqnarray}
\sigma^{noBU}_{cap}(E_{\rm c.m.})
=\frac{\pi R_b^2}{E_{\rm c.m.}}
\int_{E_{\rm c.m.}-\frac{\hbar^2\Lambda_{cr}}{2\mu R_b^2}}^{E_{\rm c.m.}}dEP^{noBU}_{cap}(E,J=0)[1-\frac{4(E_{\rm c.m.}-E)}{\mu\omega_b^2  R_b^2}]
\label{noBU}
\end{eqnarray}
in the absence of the breakup
process (Fig.~5, dotted line) by using the following formula for the capture
probability in this case~\cite{Nash}:
\begin{eqnarray}
P^{noBU}_{cap}(E_{\rm c.m.},J=0)=1-\frac{P_{qe}(E_{\rm c.m.},J=0)}{1-P_{BU}(E_{\rm c.m.},J=0)}.
\label{P_noBU}
\end{eqnarray}
By employing the  measured excitation functions $P_{qe}$ and
$P_{BU}$ at the backward angle~\cite{Li6Pb}, Eqs.~(13),~(17), and the
formula
\begin{eqnarray}
<P_{BU}>(E_{\rm c.m.})=1-\frac{\sigma_{cap}(E_{\rm c.m.})}{\sigma^{noBU}_{cap}(E_{\rm c.m.})},
\label{P_BU}
\end{eqnarray}
we  extract the mean breakup  probability $<P_{BU}>(E_{\rm c.m.})$
averaged over all  partial waves $J$~(Fig.~6).
\begin{figure}[htb]
\centering
\includegraphics[scale=0.74]{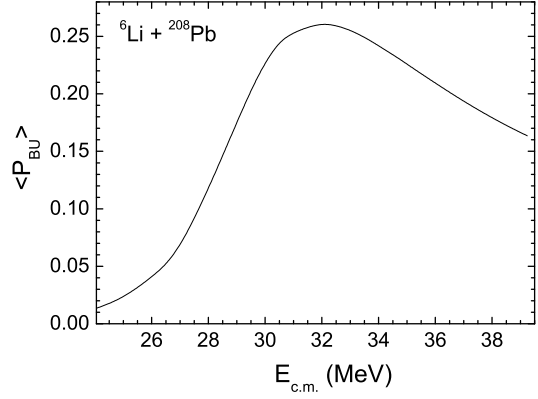}
\caption{
The extracted mean breakup probability $<P_{BU}>(E_{\rm c.m.})$ [Eq.~(19)]
as a function of bombarding energy $E_{\rm c.m.}$
for the $^{6}$Li+$^{208}$Pb reaction.
The used experimental quasi-elastic  backscattering  and quasi-elastic  backscattering  plus breakup
at the backward angle data
are from Ref.~\protect\cite{Li6Pb}.
}
\label{6_fig}
\end{figure}
The value
of $<P_{BU}>$ has a maximum at $E_{\rm c.m.}-V_b\approx 4$ MeV ($<P_{BU}>$=0.26) and slightly (sharply)
decreases with increasing (decreasing) $E_{\rm c.m.}$. The experimental
breakup excitation function at  backward angle
has the similar energy behavior~\cite{Li6Pb}.
By comparing  the calculated capture cross sections in the
absence of breakup and experimental capture (complete fusion)
data, the opposite energy trend is found in Ref.~\cite{Nash}, where
$<P_{BU}>$ has a minimum at $E_{\rm c.m.}-V_b\approx 2$ MeV  ($<P_{BU}>$=0.34) and globally increases
in  both sides from this minimum.
It is also shown in Refs.~\cite{Nash,PRSGomes4} that there are no systematic trends of breakup
in the complete fusion
reactions with the light projectiles $^{9}$Be, $^{6,7,9}$Li, and $^{6,8}$He at
near-barrier energies.
Thus, by employing the experimental quasi-elastic
backscattering, one
can obtain the additional information about the breakup process.
\begin{figure}[htb]
\centering
\includegraphics[scale=1]{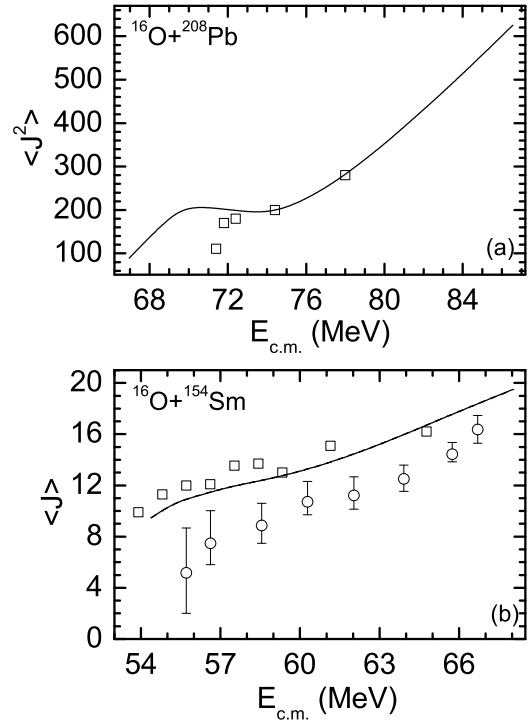}
\caption{
The extracted $<J>$ and $<J^2>$ for the  reactions $^{16}$O + $^{208}$Pb (a) and  $^{16}$O + $^{154}$Sm (b)
by employing Eqs.~(15) and  (16).
The used experimental quasi-elastic  backscattering  data are from Ref.~\protect\cite{Timmers2}.
The experimental  data of  $<J^2>$ and $<J>$ are from Refs.~\protect\cite{Vand} (open squares)
and~\protect\cite{Gil,Vand2} (open squares and circles), respectively.
}
\label{7_fig}
\end{figure}
By using the Eqs.~(15) and (16) and experimental $P_{qe}$, we extract $<J>$ and $<J^2>$ of the captured system
for the  reactions $^{16}$O + $^{154}$Sm and $^{16}$O + $^{208}$Pb, respectively (Fig.~7).
The agreements with the results of direct measurements of the $\gamma-$multiplicities in
the corresponding complete fusion reactions are quite good.
For the $^{16}$O + $^{208}$Pb reaction at sub-barrier energies,
the difference between
the extracted and experimental angular momenta is related with the deviation of the
extracted capture excitation function  from the experimental one  (see Fig.~2).

\subsection{Reaction cross sections}
\label{sec-4}
As  can be observed in Figs. 8--15, there is a good agreement between the
reaction cross sections extracted from the experimental elastic scattering at
 backward angle and from the experimental elastic scattering angular
distributions with optical potential for the reactions $^{4}$He + $^{92}$Mo,
$^{4}$He + $^{110,116}$Cd, $^{4}$He + $^{112,120}$Sn, $^{16}$O + $^{208}$Pb,
and $^{6,7}$Li + $^{64}$Zn at energies near and above the Coulomb barrier.
One can see that the used formula (\ref{3rcn_eq}) is suitable not only for
almost spherical nuclei, but also for the reactions with slightly deformed
target-nuclei. The deformation effect is effectively contained in the
experimental $P_{el}$. For very deformed nuclei, it is not possible
experimentally to separate elastic events from the low-lying inelastic
excitations. In our calculations, to obtain better agreement for the
reactions $^{16}$O+$^{208}$Pb and $^{6}$Li+$^{64}$Zn, the extracted reaction
cross sections were shifted in energy by 0.3 MeV to higher energies and 0.4
MeV to lower energies with respect to the measured experimental data,
respectively. There is no clear physical justification for the energy shift.
The most probable reason might be related with the uncertainty
associated with the elastic scattering data.
\begin{figure}[tbp]
\centering
\includegraphics[scale=0.8]{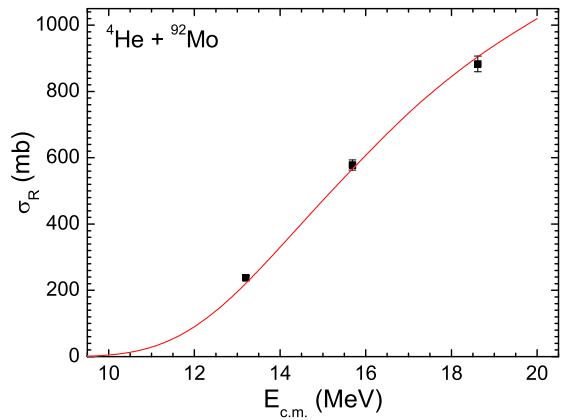}
\caption{(Color online) The extracted reaction cross sections  (solid line)
for the $^{4}$He + $^{92}$Mo reaction by employing Eq. (\protect\ref%
{3rcn_eq}). The used
experimental elastic scattering probabilities at the backward angle are from
Ref. \protect\cite{hemo}. The reaction cross sections extracted from the
experimental elastic scattering angular distribution with optical potential
are presented by   squares \protect\cite{hemo}.
}
\label{8_fig}
\end{figure}
\begin{figure}[tbp]
\centering
\includegraphics[scale=0.8]{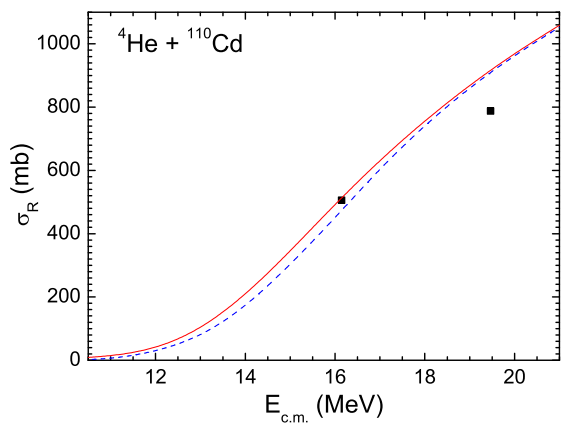}
\caption{(Color online)  The extracted reaction cross sections   (lines)
for the $^{4}$He + $^{110}$Cd reaction by employing Eq. (\protect\ref{3rcn_eq}). The used
experimental elastic scattering probabilities at the backward angle are from
Refs. \protect\cite{hecd1,hecd3} (solid line) and Ref. \protect\cite{hecd2}
(dashed line). The reaction cross sections extracted from the experimental
elastic scattering angular distribution with optical potential are presented
by   squares \protect\cite{hemo}. }
\label{9_fig}
\end{figure}
By using  Eq. (13), the capture cross sections of the reactions $^{6,7}$Li+$^{64}$Zn   can be extracted,
if one assumes that $P_{BU}=0$, since it is much smaller than $P_{qe}$.
In Figs. 14 and 15 we also show the results of our calculations for capture
cross sections of the $^{6,7}$Li+$^{64}$Zn systems, for which the fusion
process can be considered to exhaust the capture cross section. Figure 14
shows that the extracted and experimental capture cross sections are in good
agreement for the $^{6}$Li+$^{64}$Zn reaction at energies near and above the
Coulomb barrier   for the data taken in Refs. \cite{Torresi,Pietro}.
Note that the extracted capture excitation function is shifted in energy by
0.7 MeV to higher energies with respect to the experimental data.
This could be the result of different energy calibrations in the
experiments on the capture measurement and quasi-elastic scattering.
\begin{figure}[tbp]
\centering
\includegraphics[scale=0.8]{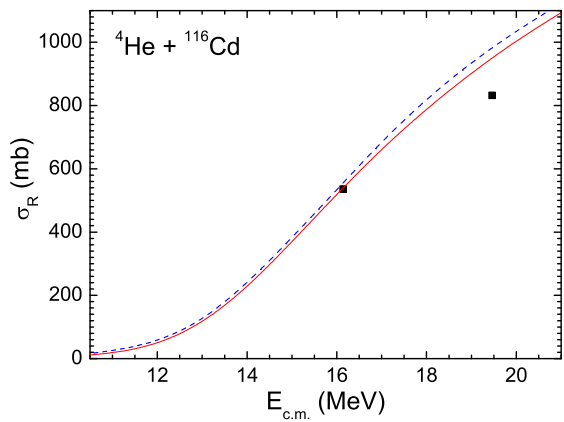}
\caption{(Color online)  The same as in Fig.~9, but for the $^{4}$He + $^{116}$Cd reaction.}
\label{10_fig}
\end{figure}
\begin{figure}[tbp]
\centering
\includegraphics[scale=0.8]{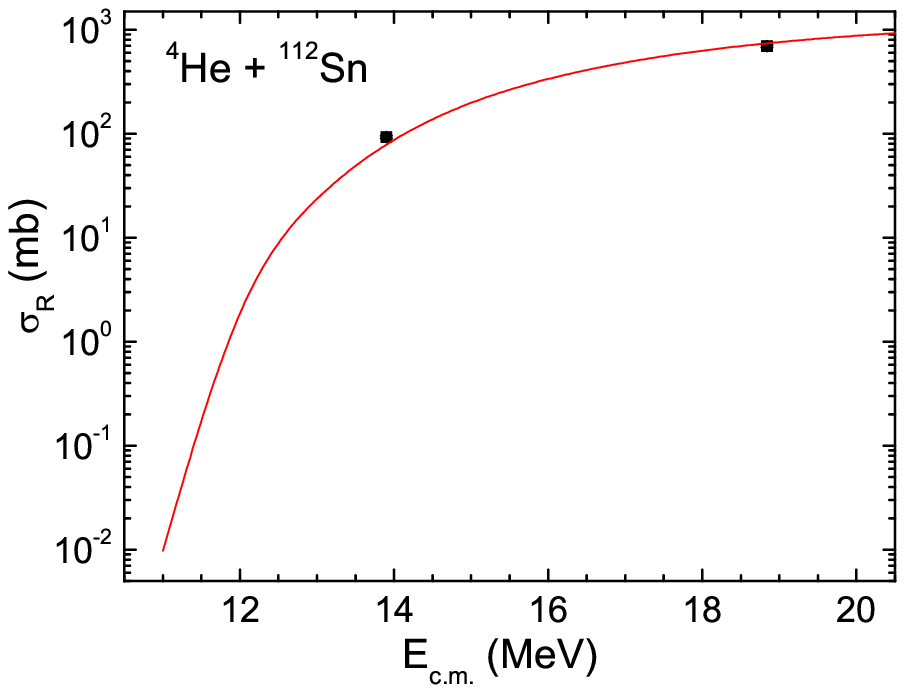}
\caption{(Color online) The extracted reaction cross sections
 (solid line) for the $^{4}$He + $^{112}$Sn reaction  by employing Eq. (\protect\ref{3rcn_eq}). The used
experimental elastic scattering probabilities at the backward angle are from
Ref. \protect\cite{hemo}. The reaction cross sections extracted from the
experimental elastic scattering angular distribution with optical potential
are presented by  squares \protect\cite{hemo}. }
\label{11_fig}
\end{figure}
\begin{figure}[tbp]
\centering
\includegraphics[scale=0.8]{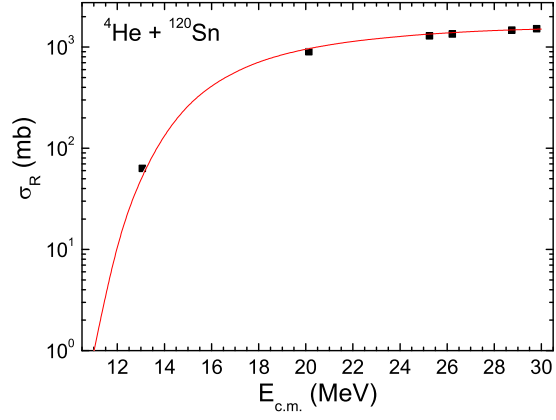}
\caption{(Color online)  The same as in Fig.~11 but for the $^{4}$He + $^{120}$Sn reaction. }
\label{12_fig}
\end{figure}
\begin{figure}[tbp]
\centering
\includegraphics[scale=0.8]{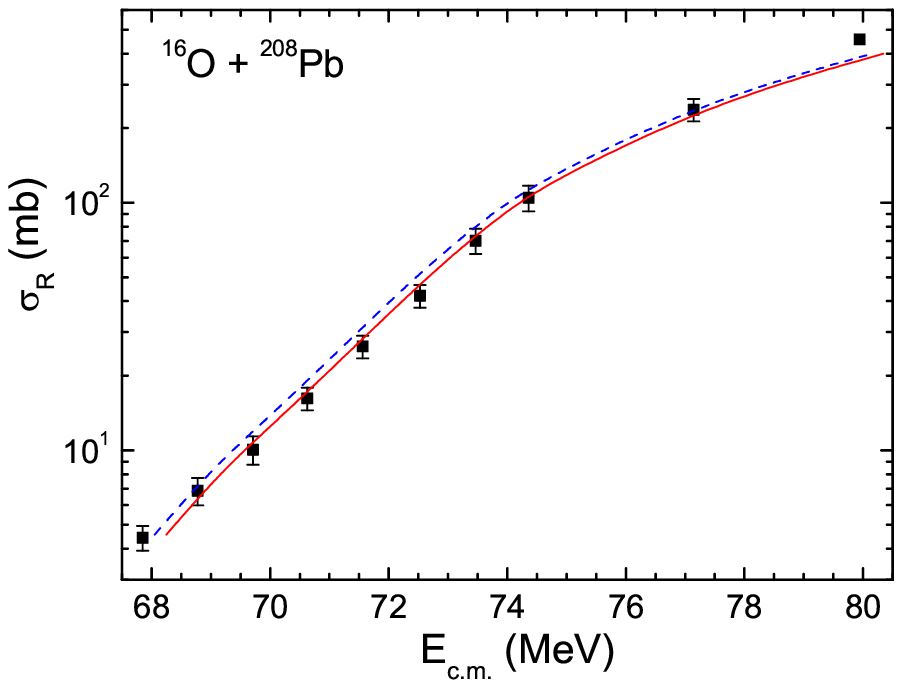}
\caption{(Color online)  The extracted reaction cross sections
(solid line) for the $^{16}$O + $^{208}$Pb reaction
by employing Eq. (\protect\ref{3rcn_eq}). The used
experimental elastic scattering probabilities at the backward angle are from
Ref.~\protect\cite{opb}. The reaction cross sections extracted from the
experimental elastic scattering angular distribution with optical potential
are presented by   squares \protect\cite{opb}.}
\label{13_fig}
\end{figure}
\begin{figure}[tbp]
\centering
\includegraphics[scale=0.8]{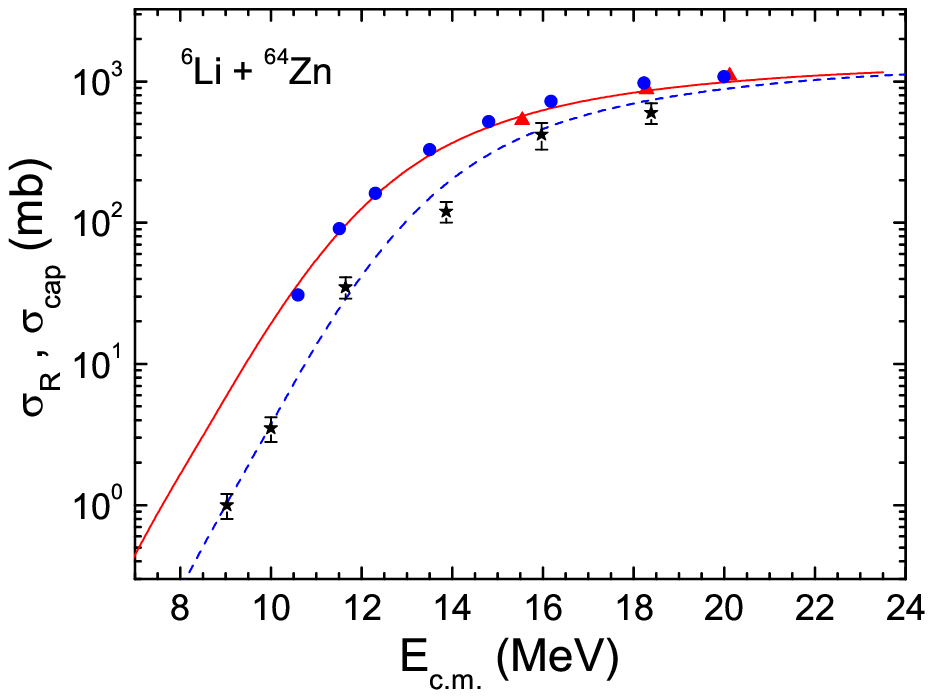}
\caption{(Color online) The extracted reaction (solid line) and capture
(dashed line) cross sections  for the $^{6}$Li + $^{64}$Zn reaction
by employing Eqs. (\protect\ref{3rcn_eq}) and (\protect\ref{3capcn_eq}). The used
experimental elastic and quasi-elastic backscattering probabilities
  are from Refs.~\protect\cite{Torresi,Pietro}. The reaction cross
sections extracted from the experimental elastic scattering angular
distribution with optical potential and capture (fusion) cross sections are
presented by   circles \protect\cite{Torresi,Pietro}, triangles \protect\cite{Gomes034,GomesPLB04}
and    stars \protect\cite{Torresi,Pietro}, respectively.
}
\label{14_fig}
\end{figure}

For the $^{7}$Li+$^{64}$Zn reaction, the $Q$-value of the one neutron
stripping transfer is positive and this process should have a reasonable
high probability to occur, whereas for the $^{6}$Li+$^{64}$Zn reaction, $Q$%
-values of neutron transfers are negative. Therefore, one might expect that
transfer cross sections for $^{7}$Li+$^{64}$Zn are larger than for $^{6}$Li+$%
^{64}$Zn.
With concern for breakup, since $^{6}$Li has a smaller threshold energy
for breakup than $^{7}$Li, one might expect that breakup cross sections for $%
^{6}$Li+$^{64}$Zn are larger than for $^{7}$Li+$^{64}$Zn. Actually, in Fig.
16, one can observe that our calculations show that
$$\sigma (^{7}\mathrm{Li}+^{64}\mathrm{Zn})>\sigma (^{6}\mathrm{Li}+^{64}\mathrm{Zn}),$$
where
$$\sigma=\sigma _{R}-\sigma _{cap}\approx \sigma _{tr}+\sigma _{in}$$
 since $\sigma_{tr}+\sigma _{in}\gg \sigma _{BU}$ for these light systems at energies
close and below the Coulomb barrier ($\sigma_{tr}$,  $\sigma_{in}$, and $\sigma _{BU}$
are the transfer,   inelastic  scattering, and breakup cross sections, respectively).
So, our present method of extracting
reaction and capture cross sections from backward elastic scattering data
allows the approximate determination of the sum of transfer and inelastic
scattering cross sections  or $\sigma _{tr}+\sigma _{in}+\sigma _{BU}$ in
systems where $P_{BU}$ cannot be neglected. For both systems investigated,
the values of these cross sections are shown to increase with $E_{\mathrm{%
c.m.}}$, reach a maximum slightly above the Coulomb barrier energy, and after,
decrease. The difference between the two curves in Fig. 16 may be considered
approximately as the difference of $\sigma _{tr}$\ between the two systems,
since $\sigma _{in}$ should be similar for both systems with the same
target, apart from the excitation of the bound excited state of $^{7}$Li.
Because $\sigma _{tr}(^{7}\mathrm{Li}+^{64}\mathrm{Zn})\gg \sigma _{tr}(^{6}%
\mathrm{Li}+^{64}\mathrm{Zn})$, one can find
$$\sigma _{tr}(^{7}\mathrm{Li}+^{64}\mathrm{Zn})\approx \sigma (^{7}\mathrm{Li}+^{64}\mathrm{Zn})-
\sigma(^{6}\mathrm{Li}+^{64}\mathrm{Zn}).$$
The maximum absolute value of the
transfer cross section $\sigma _{tr}$ at energies near the Coulomb barrier
is about 30 mb. Figure 16 also shows that the difference between transfer cross
sections for $^{7}$Li+$^{64}$Zn and $^{6}$Li+$^{64}$Zn are much more important than the possible
larger  $\sigma _{BU}$ for $^{6}$Li than for $^{7}$Li.
\begin{figure}[tbp]
\centering
\includegraphics[scale=0.8]{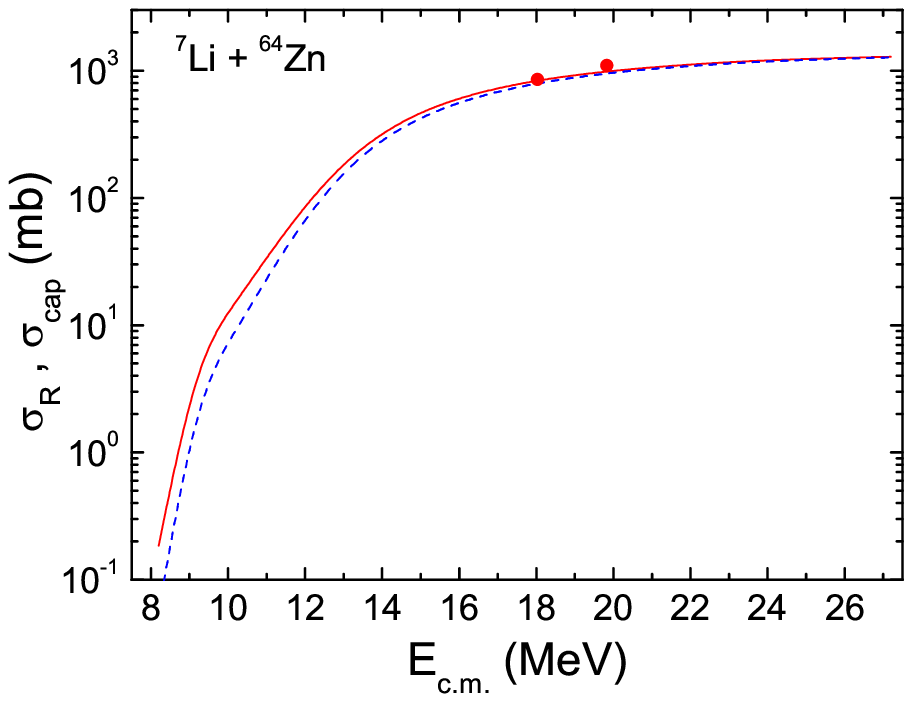}
\caption{(Color online) The same as in Fig. 14, but for the $^{7}$Li + $^{64}$Zn reaction. The reaction cross
sections extracted from the experimental elastic scattering angular
distribution with optical potential   are
presented by   circles   \protect\cite{Gomes034,GomesPLB04}.
}
\label{15_fig}
\end{figure}
\begin{figure}[tbp]
\centering
\includegraphics[scale=0.8]{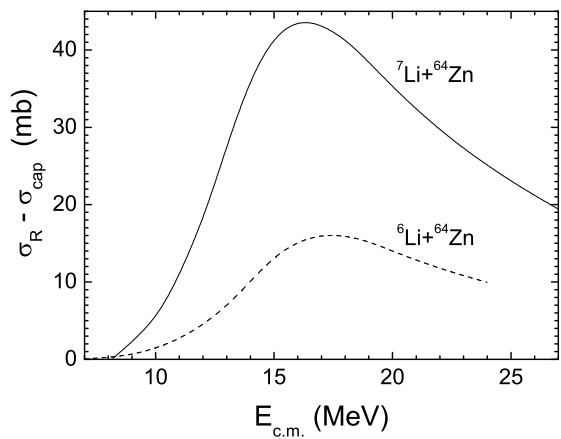}
\caption{The extracted $\protect\sigma_R - \protect\sigma_{cap}$ for the
reactions $^{6}$Li + $^{64}$Zn (dashed line) and $^{7}$Li + $^{64}$Zn (solid
line).}
\label{16_fig}
\end{figure}

\section{Summary}
We propose a new and very simple ways to determine reaction and capture (fusion) cross sections,
through the relation  (\ref{3rcn_eq})   between the elastic backscattering excitation
function and reaction cross section  and through the relation (\ref{3capcn_eq})
between the quasi-elastic scattering excitation
function at the backward angle and capture cross section. We show, for several
systems, that these methods work  well and that the elastic and quasi-elastic backscattering
technique could be used as an important and simple tools in the study of the
reaction and capture cross sections in the reactions with toughly and weakly bound nuclei.
The extraction of reaction (capture)  cross sections from the
elastic (quasi-elastic)  backscattering   is possible with reasonable
uncertainties as long as the deviation between the elastic (quasi-elastic) scattering cross
section and the Rutherford cross section exceeds the experimental
uncertainties significantly. By employing the  quasi-elastic backscattering data, one can
extract the moments of the angular momentum of the captured system.
The behavior of the transfer plus inelastic
excitation function extracted from the experimental probabilities of the
elastic and quasi-elastic scatterings at the backward angle also was  shown.

\vspace*{1cm}

We thank H.Q.~Zhang for fruitful discussions and suggestions.
We are grateful to G.~Kiss, R.~Lichtenth\"{a}ler, C.J.~Lin, P.~Mohr, E.~Piasecki,  M.~Zadro,  and H.Q.~Zhang
for providing us the experimental data.
P.R.S.G. acknowledges the partial financial support from CNPq and FAPERJ.
This work was supported by DFG, NSFC, RFBR, and JINR grants.
The IN2P3(France)-JINR(Dubna) and Polish - JINR(Dubna) Cooperation Programmes
are gratefully acknowledged.\newline
%
%
%

\end{document}